\documentclass[12pt]{article}
\usepackage{ascmac}
\usepackage{ulem}
\usepackage{latexsym}
\usepackage{amsmath}
\usepackage[top=3.5cm,bottom=3.5cm,left=2cm,right=2cm]{geometry}
\usepackage{pifont}          
\usepackage{amsmath,amssymb}
\usepackage[pdftex]{graphicx,color}
\usepackage{geometry}
\usepackage{caption}
\usepackage{ulem}
\usepackage{cancel}
\newcommand{\nn}{\nonumber\\}

\newcommand{\h}{\hspace}

\newcommand{\pa}{\partial}
\newcommand{\Slash}[1]{{\ooalign{\hfil/\hfil\crcr$#1$}}}

\newcommand{\be}{\begin{equation}}
\newcommand{\e}{\end{equation}}
\newcommand{\aln}[1]{\begin{align}#1\end{align}}

\begin{document}

\title{
\vbox{
\baselineskip 14pt
\hfill \hbox{\normalsize KUNS-2699
}} \vskip 1cm
\bf \Large  Stability of the Matrix Model in Operator Interpretation
\vskip 0.5cm
}
\author{
Katsuta~Sakai\thanks{E-mail: \tt katsutas@gauge.scphys.kyoto-u.ac.jp} 
\bigskip\\
\it \normalsize
 Department of Physics, Kyoto University, Kyoto 606-8502, Japan\\
\smallskip
}
\date{\empty}

\maketitle

\abstract{\normalsize
The IIB matrix model is one of the candidates for nonperturbative formulation of string theory, and it is believed that the model contains gravitational degrees of freedom in some manner. 
In some preceding works, it was proposed that the matrix model describes the curved space where the matrices represent differential operators that are defined on a principal bundle.  
In this paper, we study the dynamics of the model in this interpretation, and point out the necessity of the principal bundle from the viewpoint of the stability and diffeomorphism invariance.   
We also compute the one-loop correction which possibly yields a mass term for each field due to the principal bundle. 
We find that the correction does generate some mass terms with the supersymmetry broken, while fields in the original IIB matrix model remain massless.  
The positivity is not violated as long as the number of bosonic degrees of freedom is larger than the fermionic counterpart. The generation of mass terms means that the new mass scale emerges through the loop correction. 
}
\newpage

\section{Introduction and Brief Review}\label{sec:intro}
Although superstring theory is one of the candidates that include quantum gravity, its perturbative formulation seems to lack predictability in the sense that there is no criteria about the favored choice of the background or the vacuum. 
Therefore, we need its non-perturbative formulation or a new framework that includes gravity in a background manner. 
Among the various candidates, the IIB matrix model \cite{Ishibashi:1996xs,Aoki:1998bq} is promising because it can be the constructive formulation of type IIB string theory. 
Its action is given by 
\be S_{\text{IIB}}^{}=-\frac{1}{g^2}\text{Tr}\left(\frac{1}{4}[\hat{X}_a,\hat{X}_b][\hat{X}_c,\hat{X}_d]\eta^{ac}\eta^{bd} +\frac{1}{2}\bar{\hat{\Psi}}\Gamma^a[\hat{X}_b,\hat{\Psi}]\right),
\label{eq:IIB action}
\e
where $a$, $b$, $c$, and $d$ are the ten dimensional Lorentz indices. 
$\hat{X}^a$ and $\hat{\Psi}$ are a ten dimensional vector and a Majorana spinor respectively, each component of which are also $N\times N$ hermitian matrices. 
Eq.(\ref{eq:IIB action}) has the following manifest $SO(9,1)$ and $U(N)$ invariances: 
\aln{ &\delta \hat{X}_a=i\epsilon^{cd}(\hat{{\cal{O}}}_{cd}^{})_a^{~b}\hat{X}_b,\ \hat{{\cal{O}}}_{cd}^{}:D\text{-dimensional Lorentz generator}
\\ 
    &\delta \hat{X}_a=i[\hat{X}_a,\Lambda],\ \Lambda: N\times N\ \text{Hermitian matrix}\label{eq: U(N) symmetry}
}
In the theory, the Wilson loops are identified with asymptotic string states in the continuum limit \cite{Fukuma:1997en,Hamada:1997dt}, and their Schwinger-Dyson equation (loop equation) represents the time evolution of strings, including their splitting and joining. 
In addition to the original IIB matrix model, many matrix models have been investigated such as its bosonic part alone and its deformation by adding new terms or matters. 
In spite of these progresses, the meaning of matrix is not yet clear, and many interpretations have been proposed so far \cite{Aoki:1998vn,Iso:1999xs,Imai:2003vr,Imai:2003jb,Kaneko:2005pw,Kim:2011cr,Asano:2012mn,Kawai:2013wwa,Hanada:2005vr,Kawai:2007zz,Kawana:2016tkw,Hamada:2015dja}.
Among them, the operator interpretation of the matrix model \cite{Hanada:2005vr,Kawai:2007zz} is quite interesting and promising because curved spaces can be described in the diffeomorphism-covariant manner. 
In this interpretation, we interpret the large-$N$ matrices $\hat{X}_a$'s as linear operators which act on smooth functions on a given ($D\leq 10$ dimensional) manifold ${\cal{M}}$:
\be \hat{X}^a\in End(C^\infty({\cal{M}}))
\label{eq:derivative interpretation}
\e
whose operation can be explicitly written as 
\aln{ (\hat{X}_af)(x)&=\int d^D y\ X_a(x,y)f(y)\nonumber
\\
&=\left(S_a(x)+\frac{i}{2}[A_a^{~\mu}(x),\bigtriangledown_\mu^{}]_++\frac{i^2}{2}[b_a^{~\mu\nu}(x),\bigtriangledown_\mu^{}\bigtriangledown_\nu^{}]_+
+\cdots\right)f(x)\nonumber
\\
&\h{4cm} \text{for } ^\forall f(x)\in C^{\infty}({\cal{M}}),\label{eq: linear operator}
}
where we have expanded the kernel $X_a(x,y)$ by infinite local fields $\{S_a(x),A_a^{~\mu}(x),\cdots\}$ and the covariant derivative $\bigtriangledown$ on ${\cal{M}}$, 
and $[~,~]_+$ represents anti-commutator to realize hermiticity. 
In particular, if we take an ansatz 
\be \hat{X}_a^{}=i{e_a^{}}^\mu(x)\bigtriangledown_\mu^{}:=i\bigtriangledown_a^{},~~\hat{\Psi}=0
\label{eq: classical solution 1}
\e
as a classical solution where ${e_a^{}}^\mu(x)$ is the vielbein, we can show that the equation of motion (EOM) of the matrix model is equivalent to the Einstein equation. 
In fact, by substituting Eq.(\ref{eq: classical solution 1}) to the EOM
\be [\hat{X}^a,[\hat{X}_a^{},\hat{X}_b^{}]]=0,
\label{eq: eom 1}
\e
we obtain
\aln{ [\bigtriangledown^a,\left[\bigtriangledown_a^{},\bigtriangledown_b^{}\right]]&=[\bigtriangledown^a,{R_{ab}^{}}^{cd}\times\hat{{\cal{O}}}_{cd}^{}]\nonumber
\\
&=(\bigtriangledown^{a}{R_{ab}^{}}^{cd})~\hat{{\cal{O}}}_{cd}^{}-{R_b^{}}^a\bigtriangledown_{a}^{}=0.\nonumber
\\
\therefore &\bigtriangledown^{a}{R_{ab}^{}}^{cd}=0,\ {R_{b}^{}}^{a}=0,
}
where the first equation follows from the second one by using the Bianchi identity. 
In this sense, we can actually describe curved spaces by matrices $\hat{X}_a^{}$'s in the large-$N$ limit. 
It is important to note that we have treated the indices $a,b,\cdots$ in the above 
discussion as local Lorentz indices. Throughout this paper we take the same treatment.

Furthermore, we can also see the diffeomorphism invariance within the original $U(N)$ symmetry by choosing
\be 
\Lambda=\frac{i}{2}[\lambda^\mu(x),\partial_\mu^{}]_+
\e
in Eq.(\ref{eq: U(N) symmetry}). 
For example, $S_a(x)$ and $A_a^{~\mu}(x)$ transform as 
\aln{ \delta S_a(x)&=i\left[\Lambda,S_{a}\right]=\lambda^\mu(x)\partial_\mu^{}S_{a}(x),\nn 
\delta A_a^{~\mu}(x)&=-\lambda^\nu(x)\partial_\nu^{}A_a^{~\mu}(x)+(\partial_\nu^{}\lambda^\mu(x))A_a^{~\nu}(x),
}
which are the usual diffeomorphism transformations in the field theory. Therefore, the matrix model Eq.(\ref{eq:IIB action}) can  contain all the information of a curved manifold, and be a quantum theory of gravity. 
\\

Actually, the above discussion somewhat lacks precision. In curved spaces, the product of covariant derivatives cannot be understood as the product of matrices. 
For example, supposing $\hat{X}_a^{}=i\bigtriangledown_a^{}$ and multiplying $X_1^{}$ and $X_2^{}$, we obtain 
\be \hat{X}_1^{}\cdot\hat{X}_2^{}=-\bigtriangledown_1^{}\bigtriangledown_2^{}
=-\partial_1^{}\bigtriangledown_2^{}-\omega_1^{~2c}\bigtriangledown_c^{},
\label{eq: difficulty}
\e
where the the summation of index $c$ runs from 0 to 9. 
Thus, the right hand side of Eq.(\ref{eq: difficulty}) cannot be a simple product of two matrices $\hat{X}_1^{}$ and $\hat{X}_2^{}$. 
In order to overcome this situation, the full operator interpretation treat $\hat{X}_a^{}$'s as linear operators acting on smooth functions on the principal bundle $E_{\text{prin}}$ whose base space is ${\cal{M}}$, and fibre is $Spin(D-1,1)$ (or $Spin(D)$ in Euclidean case). 
The details are reviewed in section \ref{sec:principal}. 
Although the result of \cite{Hanada:2005vr,Kawai:2007zz} is mathematically rigorous, its fully general treatment seems a little difficult because of quite large degrees of freedom (DOF) of operators acting on $C^\infty(E_\mathrm{prin})$. 
In terms of local fields, there exist numerous infinite fields which are all massless at the classical level. 
Therefore, the stability of the model is in question, and needs to be studied or improved. 
In confronting such a situation, it is natural to consider first whether one can modify the interpretation so that it involves far smaller DOF. 
In particular, it is worth to investigate another but similar formulation where the principal bundle need not to be introduced. 
Such a formulation should be easy to analyze and control with respect to the stability. 
We expect that it can be constructed as long as the space of operators is closed as Lie algebra, at least around flat space.     
Although it cannot describe the curved spaces, we can get a feel for analyzing a dynamics in the operator interpretation. 
On the other hand, we have to study the original formulation in detail. 
When we discuss the stability, the primal issue is whether the fields become tachyonic due to loop corrections. 
If there is no tachyonic field found, the model can be stable in spite of the numerous DOF.   
\\

This paper is organized as follows. 
First, we attempt to construct the new minimal operator interpretation in the flat space. 
There the operators act on $End(C^\infty(\mathbb{R}^d))$, and the principal bundle is not introduced. 
Although the local Lorentz flame needs to be fixed by hand and curved spaces seems no longer to be described, such treatment is still algebraically consistent. 
In this case one has to pose an additional condition on the space of operators, which breaks general diffeomorphism invariance. 
Then, in section \ref{sec:principal} and \ref{sec:one-loop}, we consider the original operator interpretation on a principal bundle. 
In section \ref{sec:principal}, we repeat the same analysis as in section \ref{sec:minimum} and see that no additional condition is required. 
This essentially means the nessesity of the principal bundle for realizing the stability and diffeomorphism invariance. 
The latter is often needed for a theory to contain the gravitational DOF. 
Next, in section \ref{sec:one-loop}, we calculate the one-loop correction to the spectrum of fluctuation. 
We see the induced mass terms for each field does not violate positivity of the model, as long as the number of bosonic degrees of freedom is larger than the fermionic counterpart. 
Finally, in section \ref{sec:summary}, we give summary and discussion. 
Throughout this paper, we focus on the IIB matrix model with the Euclidean signature, and first analyze the bosonic part of the model.

\section{Minimum operator interpretation of Matrix Model}\label{sec:minimum}
As we mentioned in Introduction, general components of $End(C^\infty({\cal{M}}))$ cannot be understood as matrices because they are not generally closed under the multiplication or the commutator $[\ ,\ ]$. 
However, we can actually construct a set of operators which are closed under those operations by restricting $End(C^\infty({\cal{M}}))$. 
Among various possibilities, the simplest one is the set of the first order differential operators: 
\be \hat{X}_a^{}\in {\cal{A}}=\left\{f^\mu(\hat{x})\hat{p}_\mu^{}+g^{}(\hat{x})~~\Big{|}~~\hat{x}^\mu f(x)=x^\mu f(x),\ \hat{p}_\mu^{}f(x)=
-i\frac{\partial f}{\partial x^\mu},\ f^\mu,g\in C^\infty({\cal{M}})
\right\}.
\label{eq: set of op}
\e 
One can easily check that ${\cal{A}}$ is closed under the multiplication and the commutator. 
${\cal{A}}$ can also be understood as a set of quantum mechanical operators constructed by $\hat{x}^\mu$ and $\hat{p}_\mu^{}=-i\partial/\partial\hat{x}^\mu$. 
In the following discussion, we consider the semi-classical limit of those operators:
\aln{(\hat{x}^\mu,\hat{p}_\mu^{})\ &\rightarrow\ (x^\mu,p_\mu^{})
\\
[\hat{f},\hat{g}]\ &\rightarrow\ -i\{f,g\}=-i\sum_{\mu=1}^D\left(\frac{\partial f}{\partial p_\mu^{}}\frac{\partial g}{\partial x^\mu}
-\frac{\partial f}{\partial x^\mu}\frac{\partial g}{\partial p_\mu^{}}\right),
\label{eq: xppoisson}
\\
\text{Tr}(\cdots )\ &\rightarrow\ \int d^Dx\int d^Dp\left(\cdots \right).
}
In this limit, the EOM $[\hat{X}^b,[\hat{X}_b^{},\hat{X}_a^{}]]=0$ in the matrix model becomes 
\footnote{As a consistency check, we derive it from the action in the semi-classical limit. As long as the Poisson bracket satisfies the cyclicity condition  
\be \int d^Dx\int d^Dp\ f\{g,h\}= \int d^Dx\int d^Dp\ g\{h,f\},
\label{eq: poisson 1}
\e
we have 
\aln{ \delta S&=\frac{1}{g^2}\int d^Dx\int d^Dp\ \{X^a,X^b\}\{\delta X_a^{},X_b^{}\}=\frac{1}{g^2}\int d^Dx\int d^Dp\ \delta X_a^{}\{X_b^{},\{X^a,X^b\}\}\nonumber
}
which coincides with the semi-classical limit of Eq.(\ref{eq: eom 1}). 
We can easily check Eq.(\ref{eq: poisson 1}) by assuming that the integral of the total derivative terms vanish.
}
\be \{X^b,\{X_b^{},X_a^{}\}\}=0.
\label{eq: classical eom}
\e
The simplest solution of this equation is $X_a^{}=\delta_a^\mu p_\mu^{}$, and this corresponds to the flat spacetime as does in other interpretations. 
Now, let us consider the fluctuation around this solution:
\aln{ X_a^{}&=\delta_a^{\mu}p_\mu^{}+A_a^{}(x,p)\nonumber
\\
&=\delta_a^{\mu}p_\mu^{}+f_a^{}(x)+{f_a^{}}^\mu(x)p_\mu^{}\nonumber
\\
&
\equiv {e_a^{}}^\mu(x)p_\mu^{}+f_a^{}(x),\ {e_a^{}}^\mu(x)=\delta_a^{\mu}+{f_a^{}}^\mu(x).
\label{eq: fluctuation}
}
In the semi-classical limit, the bosonic part of the original IIB action becomes
\aln{
S_{\text{IIB}}^{}&=-\frac{1}{4g^2}\text{Tr}\left([\hat{p}_a^{}+A_a^{}(\hat{x},\hat{p}),\hat{p}_{b}^{}+A_{b}^{}(\hat{x},\hat{p})]^2\right)
=
\frac{1}{4g^2}\int d^Dx\int d^Dp\  G_{ab}^{}(x,p)G^{ab}(x,p),
\label{eq: classical action}
}
where
\be G_{ab}^{}(x,p)=\{X_a^{},X_b^{}\}=\delta_a^{\mu}\partial_\mu^{}A_b^{}(x,p)-\delta_b^{\mu}\partial_\mu^{}A_a^{}(x,p)
+\{A_a^{},A_b^{}\}.
\e
Furthermore, the original $U(N)$ transformation Eq.(\ref{eq: U(N) symmetry}) now becomes 
\footnote{
This gauge symmety does exist as long as the Poisson bracket satisfies the Jacobi identity 
\aln{
\{f,\{g,h\}\}+\{g,\{h,f\}\}+\{h,\{f,g\}\}=0. 
\label{eq: jacobi}
}
The Poisson bracket in this section (\ref{eq: xppoisson}) satisfies it. 
}
\aln{ \delta A_a^{}(x,p)&=\{\delta_a^\mu p_\mu^{}+A_a^{}(x,p),\Lambda(x,p)\}\nn
&=-\delta_a^\mu\partial_\mu^{}\Lambda (x,p)+\{A_a^{}(x,p),\Lambda(x,p)\},  \label{eq: classical trans}
}  
which leads to the transformation of each fields as
\aln{\delta f_a^{}(x)&={e_a^{}}^\mu(x)\partial_\mu^{}\lambda(x)-(\partial_\mu^{}f_a^{}(x))\lambda^\mu(x),\nn 
\delta {e_a^{}}^{\mu}(x)&={e_a^{}}^{\nu}(x)\partial_\nu^{} \lambda^\mu-(\partial_\nu^{}{e_a^{}}^{\mu}(x))\lambda^\nu(x),
\label{eq: each trans}
}
where we have also expanded $\Lambda(x,p)$ as $\lambda(x)+p_\mu^{}\lambda^\mu(x)$. 
Here, the first term in $\delta f_a^{}$ corresponds to the gauge transformation, and the other terms correspond to the diffeomorphism transformations. 
In particular, it is surprising that we have obtained the correct transformation low of ${e_a^{}}^\mu(x)$ as a vielbein field. 
Note that the action Eq.(\ref{eq: classical action}) is, of course, invariant under Eq.(\ref{eq: classical trans}) or Eq.(\ref{eq: each trans}) up to a total derivative term:
\be \delta S_{\text{IIB}}^{}=\frac{1}{8g^2}\int d^Dx\int d^Dp\  \{G_{ab}^{},\Lambda\}G^{ab}.
\e
Therefore, the EOM Eq.(\ref{eq: classical eom}) is also invariant under Eq.(\ref{eq: classical trans}) or Eq.(\ref{eq: each trans}). 

Let us now consider the dynamics of the fluctuations at the classical level.  
By substituting Eq.(\ref{eq: fluctuation}) to Eq.(\ref{eq: classical eom}), we obtain 
\aln{\bar{\Box}_{ab}^{}f^b+\partial_\mu^{}f^b\left(\bar{\partial}_a^{}{e_b^{}}^\mu-\bar{\partial}_b^{}{e_a^{}}^\mu\right)
+p_\mu^{}\left[
\bar{\Box}_{ab}^{}e^{b\mu}+(\partial_\nu^{}e^{b\mu})\left(\bar{\partial}_a^{}{e_b^{}}^\nu-\bar{\partial}_b^{}{e_a^{}}^\nu\right)
\right]=0,
\label{eq: eom fluctuation}
}
where
\aln{
\bar{\Box}_{ab}^{}&=\bar{\partial}_c^{}\cdot \bar{\partial}^c \delta_{ab}^{}-\bar{\partial}_a^{}\cdot \bar{\partial}_b^{}, 
\\
\bar{\partial}_a^{}&={e_a^{}}^\mu\partial_\mu^{}.
}
Note that $\bar{\partial}_a^{}$'s do not commute each other because of ${e_a^{}}^\mu$. 
Eq.(\ref{eq: eom fluctuation}) holds for arbitrary $p_\mu^{}$, so it is equivalent to the following two equations
\be \begin{cases}
\bar{\Box}_{ab}^{}f^b+\partial_\mu^{}f^b\left(\bar{\partial}_a^{}{e_b^{}}^\mu-\bar{\partial}_b^{}{e_a^{}}^\mu\right)=0,
\\
\bar{\Box}_{ab}^{}e^{b\mu}+(\partial_\nu^{}e^{b\mu})\left(\bar{\partial}_a^{}{e_b^{}}^\nu-\bar{\partial}_b^{}{e_a^{}}^\nu\right)=0.
\end{cases}
\e
From these equations, one can see that there is no suitable $D$-dimensional action that produces those EOMs because the second one has no $f_a^{}$ dependence.  
This fact means that we should not expand $X_a^{}(x,p)$ by $p_\mu^{}$ at the action level, and that we should treat the stationary point of the action Eq.(\ref{eq: classical action}) 
with respect to the matrices $X_a^{}(x,p)$.

Now let us study how many DOF remain at the liberalized (free) level:  
\be \begin{cases}
\left(\Box \delta_{ab}^{}-\partial_a^{}\partial_b^{}\right)f^b
=0,\ \delta f_a^{}=\partial_a^{}\lambda-(\partial_\mu^{}f_a^{})\lambda^\mu,
\\
\left(\Box \delta_{ab}^{}-\partial_a^{}\partial_b^{}\right)f^{b\mu}
=0,\ 
\delta {f_a^{}}^\mu=(\delta_a^{\nu}+f_a^{\nu})\partial_\nu^{} \lambda^\mu-(\partial_\nu^{}{f_a^{}}^{\mu})\lambda^\nu,
\end{cases} 
\e
where $\partial_a=\delta_a^\mu\partial_\mu$ and $\Box=\partial^\mu\partial_\mu^{}$ are the ordinary differential and d'Alembert operator, respectively. 
As for $f_a^{}$, this is completely the same as the ordinary gauge field. Thus, by choosing the Lorentz gauge, its EOM, gauge condition, and residual symmetry are given by 
\be \Box f_a^{}=0,\ \partial_a^{} f^a=0,\ \delta f_a^{}=\partial_a^{}\lambda,\ \Box\lambda=0,
\e
from which one can see that only two physical DOF remain. 
Next, as for ${f_a^{}}^\mu$, we can also choose a Lorentz-like gauge 
\be \partial_a^{}f^{a\mu}=0 \label{eq: Lorentz-like gauge}
\e
by using the diffeomorphism. As a result, we obtain the following  EOM, gauge condition and residual symmetry:
\be \Box {f_a^{}}^{\mu}=0,\ \partial_a^{}f^{a\mu}=0,\ \delta {f_a^{}}^{\mu}=\partial_a^{}\lambda^{\mu},\ \Box\lambda^{\mu}=0\label{eq:vielbein}
\e 
The above gauge conditions for $f^a$ and $f^{a\mu}$ are summarized with a condition 
for $A_a$: 
\aln{
\{p_a,A^a(x,p)\} = 0,
\label{eq: Adivergencefree1}
}
which fixes the original $U(N)$ gauge symmetry. 

In order to discuss physical DOF, let us move to the Fourier component $\tilde{f}_a{ }^\mu(k)$.  
In the light cone coordinate, we can choose $k=(k^+,0,0,0)$ without loss of generality. 
Then, the gauge fixing condition and the residual symmetry give  
\be \tilde{f}^{-\mu}=0,\ 
\tilde{f}^{+\mu}=0,
\e
so the remaining DOF are 
\be \tilde{f}^{i+},\ \tilde{f}^{i-},\ \tilde{f}^{ij}~~(i=1,2).  
\e
The first two are vectors, and the third one coincides with the massless states of bosonic closed string theory: graviton, Kalb-Ramond field and dilaton.  
In the presence of both $\tilde{f}^{i+}$ and $\tilde{f}^{i-}$, the theory violates positivity 
and is unstable.\footnote{
Although we are treating the Euclidean matrix model, the above analysis also applies to the Lorentzian model straightforwardly. In this sense we refers to the stability here.    
} 
However, we can eliminate $\tilde{f}^{i-}$ by assuming the following additional condition: 
\be \partial_\mu^{}f^{a\mu}=0,
\label{eq: additional condition}
\e
which leads to a condition for the diffeomorphism transformation:
\be \partial_\mu^{}\lambda^\mu(x)=0\ \Leftrightarrow \delta {f_a^{}}^a=0.
\label{eq: vpd}
\e 
The second equation means that the metric fluctuation is traceless, so 
the above transformation is the volume-preserving diffeomorphism. 
Note that the condition Eq.(\ref{eq: additional condition}) is consistent with the commutator. Consider 
\be \left\{{f_a^{}}^\mu p_\mu^{},{g_b^{}}^\mu p_\mu^{}\right\}=p_\mu^{}\left[(\partial_\nu^{}{f_a^{}}^\mu){g_b^{}}^\nu-(\partial_\nu^{}{g_b^{}}^\mu){f_a^{}}^\nu
\right]\equiv p_\mu^{}{F_{ab}^{}}^\mu, 
\e
then ${F_{ab}^{}}^\mu$ also satisfies 
\be \partial_\mu^{}{F_{ab}^{}}^\mu=0
\e
if ${f_a^{}}^\mu$ and ${g_b^{}}^\mu$ satisfy $\partial_\mu^{}{f_a^{}}^\mu=\partial_\mu^{}{g_b^{}}^\mu=0$.

\section{Operator Interpretation with Principal Bundle}\label{sec:principal}
In the previous section, we have considered the simplest possibility, namely $\hat{X}_a^{}\in {\cal{A}}$. 
The symmetry in that case is the volume-preserving diffeomorphism, not the general one (we refer to diffeomorphism which is not restricted to volume-preserving one as general diffeomorphism). 
There are a number of works to discuss the gravitational system that possesses only the volume-preserving diffeomorphism. 
It is sometimes called the unimodular gravity\cite{vanderBij:1981ym}. 
That theory is equivalent to the general relativity in many aspects and is a reasonable gravitational system. 
However, there seems to be no way to define the restriction Eq.(\ref{eq: additional condition}) in terms of matrices. 
Therefore we have to consider the result of the previous section unsatisfactory in the viewpoint of the matrix model. 
Furthermore, the local Lorentz transformation that act on the indices $a,b,\cdots$ is not given explicitly. 
Both of the problem suggest that the treatment in the previous section lacks some piece in order to contain gravitational DOF. 
On the other hand, the description proposed in \cite{Hanada:2005vr,Kawai:2007zz} includes the local Lorentz and general diffeomophism symmetries as a part of the original $U(N)$ symmetry Eq.(\ref{eq: U(N) symmetry})  
In this section we repeat the same analysis as in the previous section using this description. 
Although the result itself is trivial, it demonstrates the advantage of the principal bundle, that plays an essential role in the description. 
First, we briefly review the essential point of the principal bundle description.
The difficulty mentioned in section \ref{sec:intro} can be resolved by considering operators such as  
\aln{
\tilde{\nabla}_{(a)}\equiv R_{(a)}^{~~b}(g^{-1})\nabla_b, 
\label{principalbundleop}
}
which act on $C^\infty (E_\text{prin})$. Here, $R_{(a)}^{~~b}(g^{-1})$ is the matrix element of the vector representation of $Spin(D)$, 
and $\nabla_b$ is the usual covariant derivative. 
Note that the above operator is a scalar operator, {\it i.e.} an operator which is globally defined on the curved manifold $\mathcal{M}$ for each $(a)$. 
One can actually check this by the following way: 
Suppose that $U_i^{}$'s are local patches for ${\cal M}$, $t_{ij}(x)$ is the transition function on $U_i\cap U_j$, and $\nabla_b^{[i]}$ is the covariant derivative defined on $U_i^{}$. 
Then for the overlap region $U_i\cap U_j$ we have 
\aln{
\tilde{\nabla}_{(a)}^{[i]}&=R^{~~b}_{(a)}(g^{-1})\nabla_b^{[i]}\nn
&=R^{~~b}_{(a)}(g^{-1})
R^{~~c}_{~b}(t_{ij}(x))\nabla_c^{[j]}\nn
&=R^{~~b}_{(a)}((t_{ij}(x)g)^{-1})\nabla_b^{[j]}\nn
&=\tilde{\nabla}_{(a)}^{[j]}.
\label{gluecovder}
}
This relation involves no operation on the subscript $(a)$. 
One can also check that the operator (\ref{principalbundleop}) 
belongs to $End(C^\infty(E_\text{prin}))$. Therefore, if one interpret matrices as  
operators which have indices in parentheses   
\aln{
\hat{X}_{a} = X_{(a)}(x,i\tilde{\nabla},g,\hat{{\cal O}}),
}
their multiplication rule is the same as that for matrices. For example, if one take the operators 
as $\hat{X}_{(a)} = i\tilde{\nabla}_{(a)}$, then
\aln{
\hat{X}_{(1)}\cdot\hat{X}_{(2)} = -\tilde{\nabla}_{(1)}\tilde{\nabla}_{(2)}
}
in contrast to Eq.(\ref{eq: difficulty}). 
So, any operators constructed from $i\tilde{\nabla}$ can be understood as matrices, and they form 
a quite large set of operators in general. 

In order to consider the semi-classical limit, we need to define the Poisson bracket on the phase space of the principal bundle. 
Note that we also have the derivative operator 
$\hat{{\cal O}}_{ab}^{}$ in addition to $\hat{p}_\mu^{}$, and they 
are replaced with c-numbers in the semi-classical limit: 
\aln{
\hat{p}_\mu\rightarrow p_\mu,~~\hat{{\cal O}}_{ab}^{}\rightarrow t_{ab}^{}.
}
For the base space directions $(x,p)$, it is natural that we employ the same Poisson bracket as Eq.(\ref{eq: xppoisson}). 
Furthermore, it seems reasonable to assume that there is no 
nontrivial Poisson structure between the $(x,p)$ and $(g,t)$ directions. 
For the fibre directions $(g,t)$, we shall define the Poisson bracket naturally from the algebraic structure of $Spin(D)$. 
As a result, the nonzero components of the Poisson bracket are given by 
\aln{
&\{p_\mu,x^\nu\}=\delta_\mu^{~\nu},\nn
&\{t_{s},g_{ij}\}=i({\cal M}_{s}g)_{ij},~~\{t_{s},t_{t}\}=if_{stu}t_{u}.
\label{xpgtpoisson}
}
where $g_{ij}$ is an element of  $Spin(D)$ in fundamental representation\footnote{
Even though some constraint is posed on $\{g_{ij}\}$ so that $g\in Spin(D)$,  the Poisson bracket (\ref{xpgtpoisson}) is well-defined. 
} 
and ${\cal M}_s^{}$ is the fundamental representation of $\hat{{\cal O}_s^{}}$. 
In these expressions, the subscript $s$ represents antisymmetric double local Lorentz indices $[ab]$. 
For example, the last equation of Eq.(\ref{xpgtpoisson}) actually means 
$\{t_{ab},t_{cd}\}=i[(M_{ab})_{ce}t_{ed}+(M_{ab})_{de}t_{ce}]$, 
where $M_{ab}$ is the vector representation of $\hat{{\cal O}}_{ab}^{}$. 
Using Eq.(\ref{xpgtpoisson}), we can write the Poisson bracket of general functions on the principal bundle as 
\aln{
\{f,h\}\equiv
\frac{\pa f}{\pa p_\mu}\frac{\pa h}{\pa x^\mu}-\frac{\pa f}{\pa x^\mu}\frac{\pa h}{\pa p_\mu}
+i({\cal M}_sg)_{ij}\left(
\frac{\pa f}{\pa t_s}\frac{\pa h}{\pa g_{ij}}-\frac{\pa f}{\pa g_{ij}}\frac{\pa h}{\pa t_s}
\right)
+if_{stu}t_s\frac{\pa f}{\pa t_t}\frac{\pa h}{\pa t_u}.
\label{eq: poissonbracket}
}
This Poisson bracket satisfies the cyclicity condition Eq.(\ref{eq: poisson 1}) and the Jacobi 
identity Eq.(\ref{eq: jacobi}). 

The important point of the analysis is to take into account the factor $R_{(a)}^{~~b}$. 
For general functions 
$F_{(a)}^{}={R_{(a)}^{}}^c F_c^{}$ and $H_{(b)}={R_{(b)}^{}}^d H_d^{}$, 
their Poisson bracket becomes, 
\aln{
\{F_{(a)},H_{(b)}\} = R_{(a)}^{~~c}R_{(b)}^{~~d}\left[
\{F_c,H_d\}+i(M_sF)_c\frac{\pa H_d}{\pa t_s}-i\frac{\pa F_c}{\pa t_s}(M_s H)_d\right]. 
}
In particular, 
\aln{
\{p_{(a)},H_{(b)}\} &= R_{(a)}^{~~c}R_{(b)}^{~~d}\left[\pa_cH_d+i(M_sp)_c
\frac{\pa H_d}{\pa t_s}\right]\nn
&\equiv R_{(a)}^{~~c}R_{(b)}^{~~d}D_c H_d.
\label{eq: twistedderivative}
}
We have introduced {\it the twisted derivative} $D_c$ by the above equation. 

Let us now consider the fluctuation around the flat background 
$p_{(a)}^{}={R_{(a)}^{}}^b(g^{-1})\delta_b^{\mu} p_\mu^{}$: 
\aln{
X_{(a)} &=p_{(a)}+A_{(a)}\nn
&= R_{(a)}^{~~b}(g^{-1})\left[\delta_b^\mu p_\mu + A_b(x,p,g,t)\right]\nn
&=R_{(a)}^{~~b}(g^{-1})\left[\delta_b^\mu p_\mu +f_b(x,g)+f_b^\mu(x,g) p_\mu 
+\omega_b^{s}(x,g)t_s\right]\nn
&\equiv R_{(a)}^{~~b}(g^{-1})\left[e_b^{~\mu}(x,g)p_\mu+f_b(x,g)+\omega_b^{~s}(x,g)t_{s}\right]. 
\label{eq: fluctuation2}
}
Here, in the third and firth lines, we have restricted $A_b^{}(x,p,g,t)$ to be first order in $(p,t)$. 
This restriction is algebraically consistent because the Poisson bracket is closed among the first-order operators in $(p,t)$. 
The semi-classical limit of the original action then becomes
\aln{
S_{\text{IIB}}^{}&=
\frac{1}{4g^2}\int dq\ \{X_{(a)},X_{(b)}\}\{X^{(a)},X^{(b)}\}\nn
&=\frac{1}{4g^2}\int dq\  \tilde{G}_{ab}^{}\tilde{G}^{ab},
\label{eq: classical action2}
}
where
\aln{
\int dq &\equiv \int d^Dx\int d^Dp\int_{Spin(D)}\!\! dg\int_{\mathbb{R}^{d(d-1)/2}} \!\!dt, \\
\tilde{G}_{ab}&=D_aA_b-D_bA_a+
\{A_a,A_b\}+i(M_sA)_a\frac{\pa A_b}{\pa t_s}-i\frac{\pa A_a}{\pa t_s}(M_s A)_b.
}
Note that indices contracted outside the Poisson bracket can be replaced with ones  without parentheses due to the orthogonality of ${R_{(a)}^{}}^{b}(g^{-1})$. 
Of course this action is invariant under the gauge transformation written as 
\aln{
\delta A_{(a)} = \{\Lambda(x,p,g,t),p_{(a)}+A_{(a)}\},
}
which is equivalent to the transformation
\aln{
\delta A_{a} = -\pa_a\Lambda-i\frac{\pa \Lambda}{\pa t_s}(M_s p)_a
+\{\Lambda,A_a\}-i\frac{\pa \Lambda}{\pa t_s}(M_s A)_a^{}.
}
In terms of expanded fields in Eq.(\ref{eq: fluctuation2}), and the expanded gauge parameters
\aln{
\Lambda(x,p,g,t) = \lambda(x,g)+\lambda^\mu(x,g)p_\mu+\lambda^s(x,g)t_s,
}
the transformation lows are summarized as follows:
\begin{itemize}
\item ``$U(1)$'' gauge transformation
\aln{
\left\{
\begin{array}{lll}
\delta f_a = -e_a^{~\mu} \pa_\mu\lambda-i\omega_a^{~s}(\hat{M}_s\cdot \lambda),\\
\delta e_a^{~\mu} = 0,\\
\delta \omega_a^{~s} = 0. 
\label{eq: U(1)trsf}
\end{array}
\right.
} 
\item ``Diffeomorphism'' transformation
\aln{
\left\{
\begin{array}{lll}
\delta f_a = \lambda^\mu\pa_\mu f_a,\\
\delta e_a^{~\mu} =  -e_a^{~\nu}\pa_\nu\lambda^\mu+\lambda^\nu\pa_\nu e_a^{~\mu}
-i\omega_a^{~s}(\hat{M}_s\cdot \lambda^\mu),\\
\delta \omega_a^{~s} = \lambda^\mu\pa_\mu \omega_a^{~s}.
\end{array}
\right.
} 
\item ``Local Lorentz'' transformation
\aln{
\left\{
\begin{array}{lll}
\delta f_a = i\lambda^s(M_s\, f)_a+i\lambda^s(\hat{M}_s\cdot f_a),\\
\delta e_a^{~\mu} = i\lambda^s(M_s\, e^{~\mu})_a+i\lambda^s(\hat{M}_s\cdot e_a^{~\mu}),\\
\delta \omega_a^{~s} = -e_a^{~\mu}\pa_\mu \lambda^s+i\lambda^t f^s_{~tu}\,\omega_a^{~u}
-i\lambda^t(M_t\, \omega^{~s})_a+i\lambda^t(\hat{M}_t\cdot \omega_a^{~s})
-i\omega_a^{~t}(\hat{M}_t\cdot \lambda^s).
\label{eq: locallorentztrsf}
\end{array}
\right.
} 
\end{itemize}
Here, we have introduced an operator $\hat{M}_s$, which is defined as
\aln{
i(\hat{M}_s\cdot f)(x,p,g,t)\equiv 
\frac{\pa}{\pa \epsilon^s}f(x,p,e^{i\epsilon^t{\cal M}_t}g,t)\Big|_{\epsilon=0}.
}
Choosing the specific gauge parameters, which is independent from $g_{ij}^{}$, 
these transformations are the exact $U(1)$ gauge, diffeomorphism and local Lorentz 
transformations, respectively. 

If one expands each field according to Peter-Weyl theorem: 
\aln{
f(x,g)=\sum_{\text{r:irr.rep.}}R^{\left<\text{r}\right>j}_i(g)f^{\left<\text{r}\right>i}_j(x),
\label{eq: PW decomposition} 
}
the operation of $\hat{M}_s$ is equivalent to infinitesimal transformation 
for each representation: 
\aln{
(\hat{M}_s\cdot f)(x,p,g,t)=\sum_{\text{r:irr.rep.}}M^{\left<\text{r}\right>k}_{s~i}
R^{\left<\text{r}\right>j}_k(g)f^{\left<\text{r}\right>i}_j(x), 
}
where $i,j,k$ are identified to be local Lorentz indices. 

Now we focus on the dynamics of the system which has no $g$-dependence. 
Here we identify $\omega_a^{~s}$ to the spin connection. 
The EOM are given by 
\aln{
\{X^{(b)},\{X_{(b)},X_{(a)}\}\} = 0.
\label{eq: classical eom2}
}
We shall restrict the space of the operators as much as possible, posing on $e_a^{~\mu}$ and $\omega_a^{~s}$ the metricity condition: 
\aln{
\nabla_\mu e_a^{~\nu} = 0, 
}
and assume that $\omega_a^{~s}$ is torsion-free. 
In other words, we consider the first-order differential operators which contains only $f_a$ and $e_a^{~\mu}$ as the independent DOF. 
The space of such operators is closed with respect to the ordinary commutator. From these two condition one can deduce the following formula:
\aln{
\partial_a e_b^{~\mu}-\partial_b e_a^{~\mu}+\omega_{ab}^{~~c}e_c^{~\mu}-\omega_{ba}^{~~c}e_c^{~\mu}=e_a^{\mu}e_b^{\nu}(\Gamma_{\mu\nu}^\lambda-\Gamma_{\nu\mu}^\lambda)=0.
\label{eq: torsion-free condition}
}
The linearized EOM are then
\be \begin{cases}
\partial_bF^{ab}=0, \\
\Box f_a^{\mu}-\partial_a\partial^bf_b^{\mu}+\partial^b\omega_b^{~s}(iM_s)_a^{~\mu}-\partial^b\omega_a^{~s}(iM_s)_b^{~\mu}
-(\partial^b\omega_a^{~s}-\partial_a\omega^{b,s})(iM_s)_b^{~\mu}=0, \\
\Box \omega_a^{~s}-\partial_a\partial^b\omega_b^{~s}=0.
\end{cases}\label{eq: EOM_prin}\e
However, by using Eq.(\ref{eq: torsion-free condition}) and the explicit form of the vector representation matrices $(iM_s)_c^{~d}=(iM_{ab})_c^{~d}=\delta_{ac}\delta_b^{~d}-\delta_a^{~d}\delta_{bc}$, 
one can derive an equation:
\aln{
\Box f_a^{\mu}-\partial_a\partial^bf_b^{\mu}+\partial^b\omega_b^{~s}(iM_s)_a^{~\mu}-\partial^b\omega_a^{~s}(iM_s)_b^{~\mu}=0.
}
Therefore the second equation of Eq.(\ref{eq: EOM_prin}) falls into a simple equation
\aln{
\partial^b\omega_{ab}^{~~\mu}(e)-\partial_a\omega^{b~\mu}_{~b}(e)=0, 
}
and the third equation of Eq.(\ref{eq: EOM_prin}) is not independent from the second. 
Consequently, the EOM for the spacetime fluctuation is given by
\aln{
&\frac{1}{2}\Box h_a^{~\mu}-\frac{1}{2}(\partial_a\partial^bh_b^{\mu}+\partial^\mu\partial^bh_{ba})+\frac{1}{2}\partial_a\partial^\mu \delta^b_{~\nu}h_b^{~\nu}=0,\label{eq: EOM_metric}\\
&h_a^{~\mu}=f_a^{~\mu}+f^\mu_{~a}.
}
One can easily see that Eq.(\ref{eq: EOM_metric}) is equivalent to the usual linearized Einstein equation, through combining it with its own trace part.

Looking over the above equations, we find that the crucial point is that the dynamics of the vielbein $f_a^{~\mu}$ emerges only through the spin connection $\omega_a^{~s}$ 
in contrast to the previous section, where there is no cancellation in the explicit kinetic terms for $f_a^{~\mu}$. 
Furthermore, Eq.(\ref{eq: EOM_metric}) shows no dynamics of the antisymmetric part of $f_a^{~\mu}$. 
This is not a problem since the local Lorentz transformation should be used to make the local Lorentz frame parallel to the coordinate system, which means $f_a^{~\mu}$ is made symmetric. 
Once we set $e_a^{~\mu}=e^\mu_{~a}$, we need not pose further extra conditions, because the condition to eliminate the unstable mode Eq.(\ref{eq: additional condition}) 
is automatically satisfied through gauge-fixing condition for the diffeomorphism Eq.(\ref{eq: Lorentz-like gauge}). 
In this case, the theory is stable and the dynamical variables independent of the fibre coordinates $g$ are the $U(1)$ gauge field and the pure vielbein, only. 
Therefore we have showed that the theory is stable without any additional condition, as expected in the previous work. 
This feature suggest that the principal bundle is essential for equipping general diffeomorphism in the operator interpretation.

\section{One-loop corrections and induced mass terms}\label{sec:one-loop}
We have confirmed that the original description with the principal bundle is minimal possibility to contain gravity in the operator interpretation.  
Then it is necessary to study quantum correction to that model. 
In this section, we investigate some mass term induced by loop diagrams and see that the theory is still stable.   
We compute one-loop corrections to the action Eq.(\ref{eq: classical action2}) and its supersymmetrized version, and read off the mass term for each field.

In order to compute one-loop corrections, one should confront a problem of constructing the propagator. 
It is unclear whether one can define the propagator $1/D_a^2$ with Eq.(\ref{eq: twistedderivative}), because $D_a$ has the explicit dependence on the coordinates $(x,p)$. 
Instead of directly define $1/D_a^2$, we transform the coordinates and redefine functions as 
\aln{
t_s &\rightarrow \tilde{t}_s = t_s -ix^\mu\delta^a_{~\mu}(M_s)_a^{~b}\delta_b^{~\nu}p_\nu, \\
X(x,p,g,t)&\rightarrow X(x,p,g,\tilde{t}\,).
}
For the redefined functions with indices, the Poisson bracket Eq.(\ref{eq: poissonbracket}) changes to the following form: 
\aln{
\{F_{(a)},H_{(b)}\} &\longrightarrow \{F_{(a)},H_{(b)}\}^\prime\nn
&~~~=R_{(a)}^{~~c}R_{(b)}^{~~d}\left[
\frac{\pa F_c}{\pa p_\mu}\frac{\pa H_d}{\pa x^\mu}
-\frac{\pa F_c}{\pa x^\mu}\frac{\pa H_d}{\pa p_\mu}
+i({\cal M}_sg)_{ij}\left(
\frac{\pa F_c}{\pa \tilde{t}_s}\frac{\pa H_d}{\pa g_{ij}}-\frac{\pa F_c}{\pa g_{ij}}\frac{\pa H_d}{\pa \tilde{t}_s}
\right)\right.\nn
&~~~~~~~~~~~~~~~~~~~~+if_{stu}\tilde{t}_s\frac{\pa F_c}{\pa \tilde{t}_t}\frac{\pa H_d}{\pa \tilde{t}_u}\nn
&~~~~~~~~~~~~~~~~~~~~-i(M_sp)_\mu\frac{\pa F_c}{\pa p_\mu}\frac{\pa H_d}{\pa \tilde{t}_s}
-i(xM_s)^\mu\frac{\pa F_c}{\pa \tilde{t}_s}\frac{\pa H_d}{\pa x^\mu}\nn
&~~~~~~~~~~~~~~~~~~~~+i(M_sp)_\mu\frac{\pa F_c}{\pa \tilde{t}_s}\frac{\pa H_d}{\pa p_\mu}
+i(xM_s)^\mu\frac{\pa F_c}{\pa x^\mu}\frac{\pa H_d}{\pa \tilde{t}_s}\nn
&~~~~~~~~~~~~~~~~~~~~\left.+i(M_sF)_c\frac{\pa H_d}{\pa \tilde{t}_s}-i\frac{\pa F_c}{\pa \tilde{t}_s}(M_s H)_d
\right].
}
In particular, $D_cH_d \rightarrow \pa_cH_d$, and thus we can define the propagator for $A_a(x,p,g,\tilde{t}\,)$. 
For convenience, we will write the new coordinates $\tilde{t}_s$ as $t_s$ in the following.

\subsection{one-loop computation for the bosonic action}
Now let us compute the loop corrections to the action Eq.(\ref{eq: classical action2}) with the background field method. 
We consider the quantum fluctuation of $A_a$. Expanding Eq.(\ref{eq: classical action2}) as 
\aln{
A_a \rightarrow A_a + \phi_a, 
}
and adding the gauge-fixing terms, we obtain
\aln{
S=\int dq&\left[
\frac{1}{2}\pa_a A_b\pa^a A^b-\frac{1}{2}\pa^aA_a\pa^bA_b+R_{(a)}^{~~c}R_{(b)}^{~~d}\pa_cA_d\{A^{(a)},A^{(b)}\}^\prime+\frac{1}{4}\{A_{(a)},A_{(b)}\}^\prime\{A^{(a)},A^{(b)}\}^\prime\right.\nn
&~~+\frac{1}{2}\pa_a\phi_b\pa^a\phi^b+R_{(a)}^{~~c}R_{(b)}^{~~d}(\pa_c\phi_d-\pa_d\phi_c)\{\phi^{(a)},A^{(b)}\}^\prime
                +\frac{1}{2}R_{(a)}^{~~c}R_{(b)}^{~~d}(\pa_cA_d-\pa_dA_c)\{\phi^{(a)},\phi^{(b)}\}^\prime\nn
&~~+\frac{1}{2}\{\phi_{(a)},A_{(b)}\}^\prime\{\phi^{(a)},A^{(b)}\}^\prime-\frac{1}{2}\{\phi_{(a)},A_{(b)}\}^\prime\{\phi^{(b)},A^{(a)}\}^\prime
                +\frac{1}{2}\{\phi_{(a)},\phi_{(b)}\}^\prime\{A^{(a)},A^{(b)}\}^\prime\nn
&~~\left.-b\Box c+R_{(a)}^{~~c}\pa_cb\{A^{(a)},c\}^\prime
\right]. 
\label{eq: action-loop}
}
Here we have taken the Feynman gauge, and introduced the Faddeev-Popov ghost $c$ and anti-ghost $b$.  
Because we are interested in the induced mass terms for $A_a$, we calculate loop corrections with a condition 
\aln{
\pa_aA_b=0. 
} 
It is convenient to introduce the "momentum variables" $(k,r,h,u)$, which are dual to $(x,p,g,t)$, and an operator $(\mathbb{M}_s)_a^{~b}$, which is defined as
\aln{
(i\mathbb{M}_s\cdot A)_a(p,g,t) 
= \frac{\pa}{\pa \epsilon^s}\left[(e^{-i\epsilon^tM_t}A)_a(
e^{i\epsilon^vM_v}p,e^{i\epsilon^w{\cal M}_w}g,e^{i\frac{1}{2}\epsilon^{w'}{\cal M}_{w'}^\mathrm{adj}}t)\right]\Big|_{\epsilon=0},
\label{eq: definition of M}
} 
where ${\cal M_s^\mathrm{adj}}$ is the adjoint representation of $\mathcal{O}_s$. 
With these preparation, one can read off the propagators and the vertices from Eq.(\ref{eq: action-loop}). 
The factors needed for the calculation are bellows:  
\aln{
\left<\phi_a(k,r,h,u)\phi_b(-k,-r,-h,-u)\right> &= \frac{\delta_{ab}}{k^2},\\
\left<b(k,r,h,u)c(-k,-r,-h,-u)\right> &= \frac{1}{k^2}, 
}
\\

\begin{center}
\includegraphics[width=5cm]{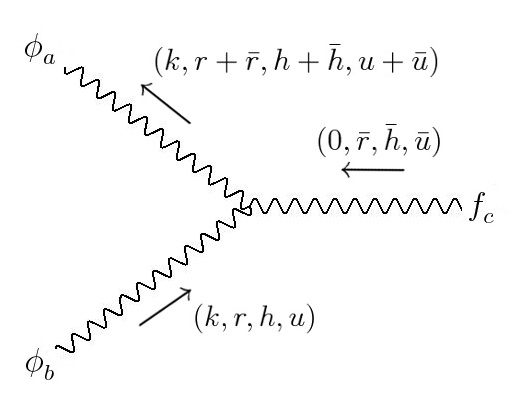}
\end{center}
\vspace{6pt}
\aln{
=i\left[(k\cdot\bar{r})\{k^b\delta^{ac}+k^a\delta^{bc}-2k^c\delta^{ab}\}
         -\{k^bu_s(\mathbb{M}_s)_e^{~c}\delta^{ae}+k^au_s(\mathbb{M}_s)_e^{~c}\delta^{be}-2(k\cdot\mathbb{M}_s)^c\delta^{ab}\}\right],
}
\\


\begin{center}
\includegraphics[width=5cm]{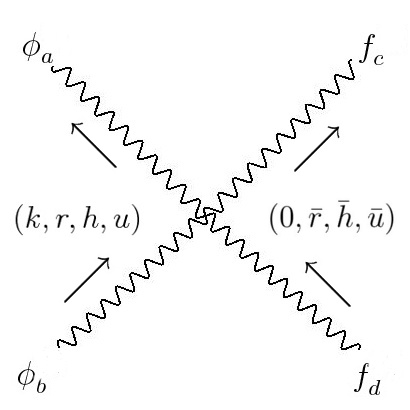}
\end{center}
\vspace{6pt}
\aln{
=&-\left[(k\cdot\bar{r})^2(2\delta^{ab}\delta^{cd}-\delta^{ad}\delta^{bc}-\delta^{ac}\delta^{bd})\right.\nn&\left.~~~~-u_su_t\{2\delta^{ab}\delta^{ef}(\mathbb{M}_s)_e^{~c}(\mathbb{M}_t)_f^{~d}
         -\delta^{ae}\delta^{bf}(\mathbb{M}_t)_e^{~d}(\mathbb{M}_s)_f^{~c}-\delta^{ae}\delta^{bf}(\mathbb{M}_t)_e^{~c}(\mathbb{M}_s)_f^{~d}\}\right],
}
\\


\begin{center}
\includegraphics[width=5cm]{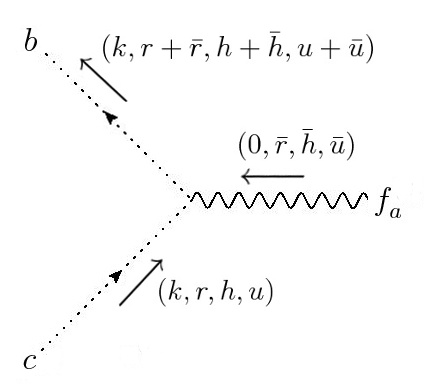}
\end{center}
\vspace{6pt}
\aln{
=i\left[-(k\cdot\bar{r})k^c+u_s(k\cdot\mathbb{M}_s)^c\right].
}
\\

%
By calculating the one-loop diagrams (Fig.\ref{fig: one-loop diagrams1}), we obtain the mass terms induced in the effective action:  
\begin{figure}
\begin{center}
\includegraphics[width=14cm]{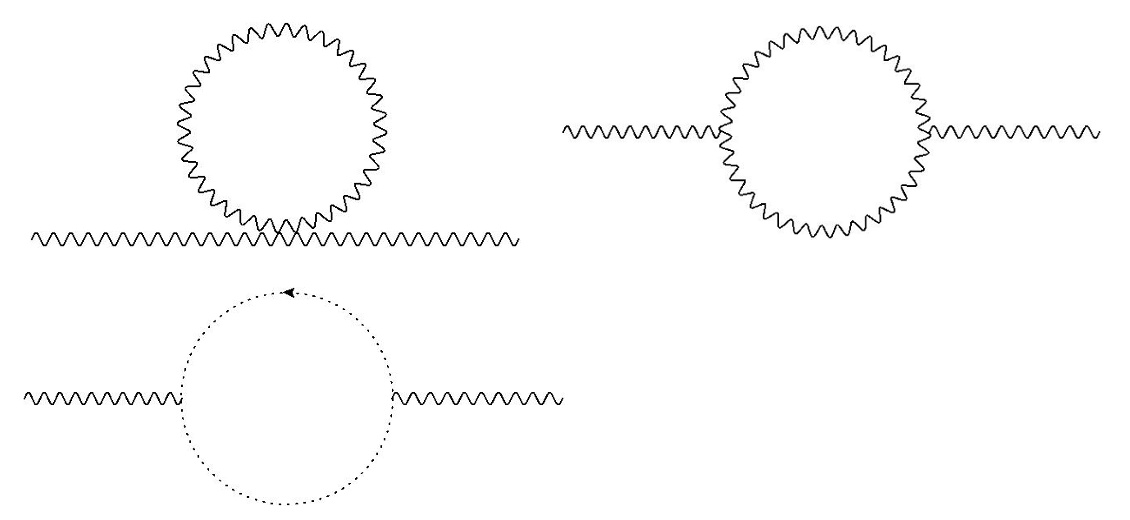}
\caption{One-loop diagrams needed for the calculation of induced mass terms.}
\label{fig: one-loop diagrams1}
\end{center}
\end{figure}
%
\aln{
\Gamma\Big|_\text{mass} &= -\frac{1}{g^2}\frac{d-2}{d+2}\int dq\left[
\alpha\left(\frac{\pa A_b}{\pa p_a}\right)^2\!-\alpha\frac{4}{d}\left(\frac{\pa A_a}{\pa p_a}\right)^2\!-\beta\frac{2(d-2)}{d^2}((\mathbb{M}_s\cdot A)_a)^2
+\gamma \left(\frac{\partial A_a}{\partial t_s}\right)^2\right],\nn
&=-\frac{1}{g^2}\frac{d-2}{d+2}\int dq\left[
\alpha\{x^{(a)}, p_{(b)}+A_{(b)}\}^2\!-\alpha\frac{4}{d}\{x^{(a)},p_{(a)}+A_{(a)}\}^2\!\right. \nn
&~~~~~~~~~~~~~~~~~~~~~~~~~~~~~~~\left.-\beta\frac{2(d-2)}{d^2}\{t_s,p_{(a)}+A_{(a)}\}^2+\gamma \{g_{ij},p_{(a)}+A_{(a)}\}^2\right],
\label{eq: mass terms}\\
\alpha&=\int \frac{d^dkd^drdhd^{d(d-1/2)}u}{(2\pi)^{d+d+d(d-1)/2}},~~\beta=\int \frac{d^dkd^drdhd^{d(d-1/2)}u}{(2\pi)^{d+d+d(d-1)/2}}\frac{u^2}{k^2},~~
\gamma\propto \int d^dk\frac{1}{k^2}\mathrm{Tr}(\mathbb{M}_s\mathbb{M}_s). 
\label{eq: divergent constants}
}
The second equality holds up to unimportant constant.
Since $\alpha$, $\beta$ and $\gamma$ are divergent, we need to take the cutoff regularization. 
Also note that the meaning of $\gamma$ is somewhat ambiguous and its numerical factor is not determined. 
However, $\mathbb{M}_s$'s can be regarded as a sort of angular momentum operators.
It is then natural to consider $\gamma$ as the sum of the eigenvalues of their Casimir operator, along with the momentum integral.   

The EOM with gauge condition is now changed as 
\aln{
\Box A_a + \alpha\frac{2(d-2)}{d+2}\left(\frac{\pa}{\pa p_b}\right)^2A_a- \alpha\frac{8(d-2)}{d(d+2)}\frac{\pa}{\pa p_a}\left(\frac{\pa A_b}{\pa p_b}\right)
-\gamma\left(\frac{\partial}{\partial t_s}\right)^2A_a~~~~~~~~~~~~\nn
- \beta\frac{4(d-2)^2}{d^2(d+2)}\left(\mathbb{M}_s\mathbb{M}_s\cdot A\right)_a=0,\label{eq: induced EOM}\\
\pa_a A^a=0.
}
To derive the above equation, one has to be careful to take variation of $((\mathbb{M}_s\cdot A)_a)^2$. 
As mentioned above, $(\mathbb{M}_s)_a^{~b}$ is equivalent to angular momentum operator, and its operation is multiplying the corresponding generator to all the indices of the field. 
Therefore it should be identified to be the derivative with respect to fibre direction, 
and one obtains in the action $((\mathbb{M}_s\cdot A)_a)^2=-A^a(\mathbb{M}_s\mathbb{M}_s\cdot A)_a$ by partial integration. 

One can analyze Eq.(\ref{eq: induced EOM}) by expanding $A_a$. 
Note that the three terms in the first line in Eq.(\ref{eq: induced EOM}) vanish since we have restricted $A_a$ to be first order in $p$ and $t_s$. 
From the last term in Eq.(\ref{eq: induced EOM}), we get mass terms for each field component. 
While $\mathbb{M}_a$ can be interpreted as the derivative with respect to the fibre coordinates, it is the generators of $Spin(D)$ and $(\mathbb{M}_s)^2$ is the Casimir operator. 
Consequently, each field arising from P-W expansion Eq.(\ref{eq: PW decomposition}) gets the positive mass squared, 
the value of which is the eigenvalue of the Casimir operator according to the representation. 
The important point here is that a field of any nontrivial representation of Lorentz group, which has implicit $g$-dependence, acquires a mass term. 
This means that vielbein fluctuation $f_a^{~\mu}(x)$ get massive as well, even though it has no explicit $g$-dependence.

\subsection{Including the fermionic sector}
The above result seems to be quite bad news for us, since there is no gravitational field when one take into account the quantum correction. 
However, the original IIB matrix model has the fermionic sector as well, and it is possible that its quantum correction drastically changes the result, as is the case in most supersymmetric theories. 
Therefore, we shall repeat the same analysis as above on the action obtained from the full IIB matrix model. 
In this subsection, we write the essence of the analysis briefly.

The action corresponding Eq.(\ref{eq: action-loop}) is now   
\aln{
S = S|_\text{bos} + \frac{1}{2}\bar{\psi}^TC\Gamma^{a}\pa_a\psi+\frac{1}{2}(\bar{\psi}^TC\Gamma^{(a)})_{(\alpha)}\{A_{(a)},\psi_{(\alpha)}\}^\prime, 
\label{eq: action-loop2}
}
where $S|_\text{bos}$ is the bosonic part Eq.(\ref{eq: action-loop}), $\Gamma^a$ is the $d$-dimensional gamma matrix and $C$ is the charge conjugation matrix. 
One can easily check that the Eq. (\ref{eq: action-loop2}) is obtained as the semi-classical limit of the IIB matrix model Eq.(\ref{eq:IIB action}).  
The new parts needed for computing loop diagrams is bellow: 
\aln{
\left<\psi_\alpha(k,r,h,u)\psi_\beta(-k,-r,-h,-u)\right> = \frac{i(\Slash{k}C^{-1})_{\alpha\beta}}{k^2}, 
}

\begin{center}
\includegraphics[width=5cm]{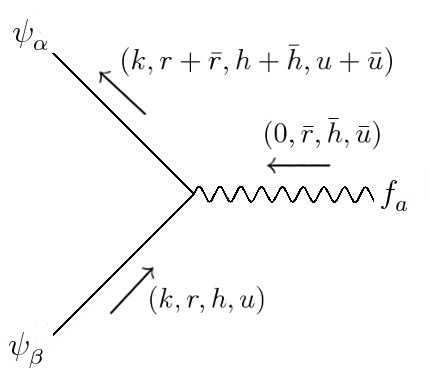}
\end{center}
\vspace{6pt}
\aln{
=i\left[(k\cdot\bar{r})\{k^b\delta^{ac}+k^a\delta^{bc}-2k^c\delta^{ab}\}
         -\{k^bu_s(\mathbb{M}_s)_e^{~c}\delta^{ae}+k^au_s(\mathbb{M}_s)_e^{~c}\delta^{be}-2(k\cdot\mathbb{M}_s)^c\delta^{ab}\}\right],
}
\\

Using them, we compute the loop diagrams containing a fermion loop (Fig.\ref{fig: one-loop diagram2}) , and obtain the induced mass terms in this case: 
\begin{figure}
\begin{center}
\includegraphics[width=6cm]{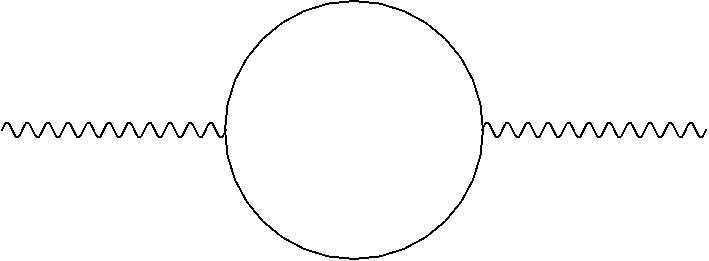}
\caption{The additional loop diagram. The loop is of Majorana fermion.}
\label{fig: one-loop diagram2}
\end{center}
\end{figure}
%
\aln{
\Gamma\Big|_\text{mass} &= -\frac{1}{g^2}\frac{1}{d+2}\left((d-2)-2^{[d/2]-\kappa}\right)\int dq\left[
\alpha\left(\frac{\pa A_b}{\pa p_a}\right)^2\!-\alpha\frac{4}{d}\left(\frac{\pa A_a}{\pa p_a}\right)^2\!-\beta\frac{2(d-2)}{d^2}((\mathbb{M}_s\cdot A)_a)^2\right], \\
\kappa&=\left\{\begin{array}{ll}
1~~(\text{for the Majorana fermion})\\
2~~(\text{for the Majorana-Weyl fermion})
\end{array}\right.. 
\label{eq: mass terms2}
}
From this induced mass term, we conclude that all of the fields, including the vielbein, remain massless in the IIB matrix model (d=10).

\section{Summary and discussion}\label{sec:summary}
In this paper, we have analyzed the stability of the matrix model in the operator interpretation, which is originally proposed in \cite{Hanada:2005vr,Kawai:2007zz}. 
We have shown that the principal bundle is essential for both the general diffeomorphism invariance and stability. 
We further have shown that the mass terms induced by loop correction do not violate the stability. 
In section 2, we have analyzed the possible minimal case, where the fundamental fields are the $U(1)$ gauge field and vielbein, along with the Kalb-Ramond and dilaton fields. 
Although the model is algebraically consistent, we have to choose the local Lorentz frame by hand and the general diffeomorphism invariance is lost by requiring an additional restriction Eq.(\ref{eq: additional condition}), which seems unnatural in terms of matrices.   
In section 3, we have introduced the principal bundle as was done in \cite{Hanada:2005vr,Kawai:2007zz}, and have seen that the model indeed contains the gravitational field. 
There, in the explicit $g$-independent sector, $f^a(x)$ is the $U(1)$ gauge field and $e_a^{~\mu}(x)$ is identified to the vielbein itself. 
On the other hand, $\omega_a^{~s}(x)$ represents the spin connection which is assumed to be written in terms of $e_a^{~\mu}$.
Finally in section 4, we have explicitly calculated the loop corrections by using the semi-classical limit, and have seen that the gauge field $f_a$ and the vielbein fluctuation $f_a^{~\mu}$ remain massless in the IIB matrix model, while they acquire mass terms in the absence of supersymmetry. In particular, the positivity of the theory is not violated when the supersymmetry is broken in the manner that the number of bosonic degrees of freedom is larger than the fermionic counterpart.

There still remain some open questions. First of all, the metricity condition does not seem to be obtained naturally from the theory, although a consistent description was obtained by assuming it. 
The condition is often crucial, because it can vary the order of EOM for fundamental variables (for example, the EOM obtained from a gravitational action with quadratic curvature is forth-order in terms of the metric but second-order in terms of the spin connection). 
It is interesting to construct the model which yields the metricity condition as EOM or consistency condition. 
Secondly, we have studied the model at linearized level in the operator interpretation. 
While one can find the gravitational DOF, {\it i.e.} the massless field which can be identified to vielbein fluctuation, it is still unclear whether it mediates the real gravitational interaction.  
In the original interpretation\cite{Ishibashi:1996xs}, the correct gravitational interaction is realized due to supersymmetry. 
Our result has the same feature as that work, from the viewpoint that supersymmetry realizes gravity.  
It is then valuable to discuss the operator interpretation at nonlinear level, and to compare scattering amplitudes with those in the original interpretation or other theories of gravity.     
Another important question is the detail of the generation of the mass scale. A new scale emerging through radiative correction is often discussed in the context of classically conformal systems. the connection between them and the matrix model is worth investigating.

\section*{Acknowledgement} 
KS thanks H. Kawai and K. Kawana for helpful discussion.  
This work is supported by Grant-in-Aid for JSPS Research Fellow Number 17J02185.



\end{document}